\documentclass[11pt]{article}

\usepackage{amsmath}
\usepackage{amscd}
\usepackage{amsfonts}
\usepackage{epsfig}
\usepackage{theorem}
\usepackage{authblk}
\usepackage[titletoc]{appendix}
\usepackage[margin=0.5in]{geometry}

\newtheorem{theorem}{Theorem}[section]

\newtheorem{lemma}{Lemma}[section]

\newtheorem{definition}{Definition}[section]

\newtheorem{proposition}{Proposition}[section]

\newtheorem{conjecture}{Conjecture}[section]

\def \C {\mathbb C}

\def \R {\mathbb R}

\def \N {\mathbb N}

\def \Z{\mathbb Z}

\title{Absorption Probabilities of Quantum Walks}

\author{Parker Kuklinski \and Mark Kon}

\begin{document}


\maketitle

\begin{abstract}
Quantum walks are known to have nontrivial interaction with absorbing boundaries. In particular, Ambainis et.\ al.\ \cite{ambainis01} showed that in the $(\Z ,C_1,H)$ quantum walk (one-dimensional Hadamard walk) an absorbing boundary partially reflects information. These authors also conjectured that the left absorption probabilities $P_n^{(1)}(1,0)$ related to the finite absorbing Hadamard walks $(\Z ,C_1,H,\{ 0,n\} )$ satisfy a linear fractional recurrence in $n$ (here $P_n(1,0)$ is the probability that a Hadamard walk particle initialized in $|1\rangle |R\rangle$ is eventually absorbed at $|0\rangle$ and not at $|n\rangle$). This result, as well as a third order linear recurrence in initial position $m$ of $P_n^{(m)}(1,0)$, was later proved by Bach and Borisov \cite{bach09} using techniques from complex analysis. In this paper we extend these results to general two state quantum walks and three-state Grover walks, while providing a partial calculation for absorption in $d$-dimensional Grover walks by a $d-1$-dimensional wall. In the one-dimensional cases, we prove partial reflection of information, a linear fractional recurrence in lattice size, and a linear recurrence in initial position.
\end{abstract}

\section{Introduction}

The quantum walk is a unitary analogue of the classical random walk. Several introductory papers have been written on quantum walks which explore their mathematical properties and their use in algorithms for quantum computers \cite{kempe03} \cite{venegas-andraca12}. The quantum walk differs from the classical random walk in many ways including its linear spreading and initial condition-dependent asymmetries. In this paper, we direct our attention to the absorption problem; that is, we wish to calculate the probability that a quantum walk particle is eventually absorbed by a collection of absorption units present in the system. By an absorption unit, we mean an element $g$ of the group $G$ that the walk takes place on such that if the quantum particle is observed at the location of the unit, then the particle is absorbed and the walk is terminated. Otherwise, we continue to alternate the quantum walk operator with these measurements until the particle is absorbed.

Solutions to the absorption problem are well known in the symmetric classical random walk \cite{spitzer13} \cite{lovasz94}. The one dimensional symmetric random walk is recurrent, meaning that a random walk particle will eventually return to its starting location (and by extension any other location on the line) with probability 1. Thus, in the semi-infinite random walk where there is a single absorbing boundary at the origin, the random walk particle will be absorbed there with probability 1. This property also holds for the two dimensional random walk but not for random walks of dimension three and higher \cite{gut12} \cite{billingsley08}. Consider also the finite absorbing random walk in which there are absorbing boundaries at positions 0 and $N$. If the particle starts at position 1, it has been shown that the probability that the particle will eventually be absorbed by the left boundary (at the origin) is equal to $1-\frac{1}{N}$. Notice that as $N\rightarrow\infty$, this probability approaches 1, so in some way the finite absorption probability limits to the semi-infinite absorption probability. These results can be generalized to asymmetric random walks, and while these walks are not recurrent, we again find that the finite absorption probabilities limit to the semi-infinite absorption probabilities.

The absorption problem for the one dimensional quantum walk is much more nuanced than the classical absorption problem. Ambainis et.\ al.\ \cite{ambainis01} first considered the absorption problem for both the semi-infinite Hadamard walk and the finite Hadamard walk. Following their notation, let $p_\infty$ be the probability that a Hadamard walk particle initialized in $|1\rangle |R\rangle$ is eventually absorbed at $|0\rangle$, and let $p_n$ be the probability that the Hadamard walk particle initialized at $|1\rangle |R\rangle$ is eventually absorbed at $|0\rangle$ if there exists another absorbing boundary at $|n\rangle$. Ambainis et.\ al.\ proved, using properties of the Catalan numbers, that $p_\infty =\frac{2}{\pi}$. They also proved that $\lim_{n\rightarrow\infty}p_n=\frac{1}{\sqrt{2}}$. These two calculations directly oppose the two key properties we highlighted about the absorption problem for the classical random walk; namely, the quantum walk is not recurrent, and the finite absorption probability does not limit to the semi-infinite absorption probability. Paradoxically, these authors showed that the presence of an absorbing boundary far away from the origin actually \emph{increases} the probability of eventual absorption at the origin.

In this same paper, Ambainis et.\ al.\ conjectured a recursion governing the finite absorption probabilities, namely $p_{n+1}=\frac{1+2p_n}{2+2p_n}$. Partial progress toward proving this result was made by Konno et.\ al.\ \cite{konno02_1} and Bach et.\ al.\ \cite{bach04} before finally being proved by Bach and Borisov \cite{bach09}. This result was obtained by using the integral representation of the Hadamard product along with a few key observations from complex analysis. These authors also cited a recursion governing the absorption probabilities based on initial position. If $p_n^{(m)}$ is the probability that the Hadamard walk particle initialized at $|m\rangle |R\rangle$ is eventually absorbed at $|0\rangle$ given an additional absorbing boundary at $|n\rangle$, then the following formula holds:
$$p_n^{(m+3)}-7p_n^{(m+2)}+7p_n^{(m+1)}-p_n^{(m)}=0.$$
These results were extended to the three-state Grover walk in Wang et.\ al.\ \cite{wang16} where it was discovered that those absorption probabilities satisfy $p_{n+1}=\frac{2+3p_n}{3+4p_n}$. Other releated papers consider the \emph{hitting time} of the quantum walk, or the average time it takes for the quantum walk particle to be absorbed \cite{yamasaki02}\cite{krovi06}\cite{montero13}.

In this paper, we extend the described results to a more general collection of quantum walks. We find that the absorption probabilities for all of these walks share three key properties; these properties are partial reflection of information at the boundary, a linear fractional recurrence in domain size, and a linear recurrence in initial position. The workflow for these calculations is consistent between examples. First, we define generating functions for which the $t^\text{th}$ coefficient of the Taylor expansion corresponds to the amplitude at the absorbing unit at time $t$. These generating functions can be explicitly computed using path counting arguments. From here, the absorption probability may be obtained by evaluating a Hadamard product of these generating functions at 1. The Hadamard product has an integral representation which for semi-infinite walks may be directly integrated. For the finite walks, the corresponding integrand can be separated by means of a partial fractions expansion. The partial fractions contain specific collections of poles that facilitate applications of the residue theorem.

In Section 2, we discuss the definitions and methods necessary to carry out our computations. In Section 3, we compute absorption probabilities for the classical random walk as a simple case. The remaining sections are devoted to computing absorption probabilities for specific variants of the quantum walk. These include the general two-state one-dimensional quantum walk in Section 4, the three-state one-dimensional Grover walk in Section 5, and the $d$-dimensional Grover walk in Section 6.

\section{Definitions and Methods}

The quantum walk has a natural construction on groups.
\begin{definition}
Let $(G,\cdot )$ be a group, let $\Sigma\subset G$ where $|\Sigma |=n$, and let $U\in U(n)$ where $U(n)$ is the set of $n\times n$ unitary matrices. The \emph{quantum walk operator} $Q:\ell ^2(G\times\Sigma )\rightarrow\ell ^2(G\times\Sigma )$ corresponding to the triple $(G,\Sigma ,U)$ may be written as $Q=T(I\otimes U)$ where for $g\in G$ and $\sigma\in\Sigma$, $T:|g\rangle |\sigma\rangle\mapsto |\sigma +g\rangle |\sigma\rangle$. We denote this correspondence as $Q\leftrightarrow (G,\Sigma ,U)$. 
\end{definition}
We refer to states $|\sigma\rangle\in\ell^2(\Sigma )$ as internal states and $|g\rangle\in\ell ^2(G)$ as position states. The pair $(G,\Sigma )$ can be thought of as an undirected Cayley graph which admits loops \cite{diestel05}. Here, the group action is defined notionally as addition for consistency. In this paper the groups will take the form $G=\Z ^n$ and the directional subsets $\Sigma\subset\Z ^n$ will either be $\Sigma =C_k:=\{ x\in\Z ^n:\lVert x\rVert =1\}$ or $\Sigma =\tilde{C_k}=C_k\cup\{ 0\}$.

We pause to make a remark about our use of braket notation. If $(g,\sigma )\in G\times\Sigma$ is an element in the underlying space, then $\{ |g\rangle |\sigma\rangle :(g,\sigma )\in G\times\Sigma\}$ is an orthonormal basis for $\ell ^2(G\times\Sigma )$. We use the braket in general to refer to an element of $\ell ^2(G\times\Sigma )$, but it will sometimes suit us to surpress the braket as will be the case in our path counting arguments which are firmly grounded in the underlying classical space. For example, if $|\psi\rangle\in\ell ^2(G\times\Sigma )$ is an element of the orthonormal basis, in the path counting arguments it will be easier to discuss $\psi\in G\times\Sigma$. We will address this distinction when necessary.

We must also define an absorption unit for quantum walks. To this end, we formally define the measurement operator. Let $b\in G\times\Sigma$. The measurement operator $\Pi ^b_\text{yes}:\ell ^2(G\times\Sigma )\rightarrow\ell ^2(G\times\Sigma )$ is a projection onto $|b\rangle$ while $\Pi _\text{no}^b$ is a projection onto the the subspace spanned by elements in $(G\times\Sigma )\backslash b$. The probabilistic interpretation of quantum mechanics dictates that if we measure a state $\psi\in\ell ^2(G\times\Sigma )$ at $|b\rangle$, the resulting state becomes $\frac{\Pi _\text{yes}^b\psi}{\lVert\Pi _\text{yes}^b\psi\rVert}$ with probability $\lVert\Pi _\text{yes}^b\psi\rVert ^2$ and $\frac{\Pi _\text{no}^b\psi}{\lVert\Pi _\text{no}^b\psi\rVert}$ with probability $\lVert\Pi _\text{no}^b\psi\rVert ^2$. If $B\subset G\times\Sigma$, let $\Pi _\text{no}^B$ be the composition of \emph{no} measurement projections for all $b\in B$. In this way, we can define the quantum walk operator for an absorbing quantum walk.
\begin{definition}
Let $Q\leftrightarrow (G,\Sigma ,U)$ be a quantum walk operator and let $B\subset G\times\Sigma$. Then we say that $\Pi _\text{no}^BQ$ is the \emph{absorbing quantum walk operator} corresponding to the ordered quadruple $(G,\Sigma ,U,B)$ and we denote this correspondence as $\Pi _\text{no}^BQ\leftrightarrow (G,\Sigma ,U,B)$.
\end{definition}
We use the \emph{no} operator in our definition because if we observe the particle somewhere in $B$, then the experiment is terminated, while if the particle is not observed in $B$ (i.e. we are in the range of $\Pi _\text{no}^B$) the experiment continues. Note that we speak of the absorption units as being elements of the classical space and not as members of the corresponding orthonormal basis. 

For an absorption problem, we wish to compute the probability $P$ that the quantum walk particle is eventually absorbed by some subset of the absorbing units $B_0\subset B$ of an absorbing quantum walk with operator $\Pi _\text{no}^B Q\leftrightarrow (G,\Sigma ,U,B)$ before it is absorbed anywhere else in $B$ (i.e. before it is absorbed in $B\backslash B_0$). Let $\psi\in\ell ^2(G\times\Sigma )$ be the initial state. Then we can express this probability as an infinite sum:
\begin{align}
P=\sum _{t=1}^\infty\left[\sum _{b\in B_0}|\langle b|Q\left(\Pi _\text{no}^B Q\right) ^{t-1}|\psi\rangle |^2\right] .\
\end{align}
This paper is devoted to computing such sums for a selection of quantum walks.

\subsection*{Path Counting and Generating Functions}

The quantum walk is a quantum Markov chain \cite{gudder14} and thus has a pathwise representation. We must first define a path and its associated amplitude and displacement.
\begin{definition}
Consider the quantum walk operator $Q\leftrightarrow (G,\Sigma ,U)$ where $U(\sigma _1,\sigma _2)$ is the transition amplitude from internal state $|\sigma _1\rangle$ to $|\sigma _2\rangle$. An \emph{$n$-path} $\gamma\in\Sigma ^{n+1}$ is represented as $\gamma =(\sigma _0,...,\sigma _n)$ and its associated \emph{displacement} is defined as $S(\gamma )=\sum _{k=1}^n \sigma _k$. Its associated \emph{amplitude} is defined as $A(\gamma )=\prod _{k=1}^n U(\sigma _{k-1},\sigma _k)$.
\end{definition}
To be consistent with Definition 2.1, we define the summation inductively by $\sum _{k=1}^n\sigma _k=\sigma _n+\sum _{k=1}^{n-1}\sigma _k$. Let $\varphi _k:\Sigma ^{n+1}\rightarrow\Sigma$ be the $k^\text{th}$ entry of a $n$-path where $0\le k\le n$. In view of the quantum walk as a quantum Markov chain, the following pathwise representation holds \cite{gudder14}
\begin{proposition}
Consider the quantum walk operator $Q\leftrightarrow (G,\Sigma ,U)$ and let us define a set of paths where $g_0,g_1\in G$ and $\sigma _0,\sigma _1\in\Sigma$:
\begin{align}
\Gamma =\{\gamma\in\Sigma ^{n+1}:\varphi _0(\gamma )=\sigma _0,\varphi _n(\gamma )=\sigma _1,S(\gamma )=g_1+ g_0^{-1}\} .\
\end{align}
Then:
\begin{align}
\langle g_1,\sigma _1|Q^n|g_0,\sigma _0\rangle =\sum _{\gamma\in\Gamma}A(\gamma ).\
\end{align}
\end{proposition}
The proof of this proposition follows from the linearity of $Q$ and an induction argument. This proposition is closely related to the path integral formulation of quantum mechanics proposed by Feynman \cite{feynman10}.

With a little care, a pathwise representation of amplitudes for absorbing quantum walks can also be written. To this end let $\theta _k:\Sigma ^{n+1}\rightarrow\Sigma ^{k+1}$ be 
the restriction of an $n$-path $\gamma =(\sigma _0,...,\sigma _n)$ to a $k$-path $\theta _k(\gamma )=(\sigma _0,...,\sigma _k)$ for $k\le n$.
\begin{proposition}
Consider the absorbing quantum walk operator $\Pi _\text{no}^BQ\leftrightarrow (G,\Sigma ,U,B)$. If we let $\Gamma$ be the set from the previous proposition, then we define a new set of paths:
\begin{align}
\Gamma '=\Gamma\cap\{\gamma\in\Sigma ^{n+1}:S(\theta _k(\gamma ))+ g_0\notin B\hspace{0.2cm}\forall k\le n\} .\
\end{align}
Then the following holds:
\begin{align}
\langle g_1,\sigma _1|\left(\Pi _\text{no}^BQ\right) ^n|g_0,\sigma _0\rangle =\sum _{\gamma\in\Gamma '}A(\gamma ).\
\end{align}
\end{proposition}
The proof of this proposition follows from proposition 2.1 and our definition of the measurement operator.

In view of the absorbing probability equation (1) from above, we consider for a complex variable $z$ generating functions $f_{b|\psi}(z)$ for an absorbing quantum walk defined as:
\begin{align}
f_{b|\psi}(z)=\sum _{t=1}^\infty \langle b|Q\left(\Pi _\text{no}^B Q\right) ^{t-1}|\psi\rangle z^t.\
\end{align}
We use the propositions above to write a pathwise representation of these generating functions:
\begin{definition}
Let $f_{b|\psi}(z)$ be a generating function for an absorbing quantum walk $(G,\Sigma ,U,B)$ where $\psi =(g_0,\sigma _0)$ and $b=(g_1,\sigma _1)$. We define the set $F_{b|\psi}^{(n)}$ of \emph{associated $n$-paths} of this generating function as:
\begin{align}
F_{b|\psi}^{(n)}=\{\gamma\in\Sigma ^{n+1}:\varphi _0(\gamma )=\sigma _0,\varphi _n(\gamma )=\sigma _1,S(\gamma )=g_1+ g_0^{-1},S(\theta _k(\gamma ))+ g_0\notin B\hspace{0.2cm}\forall k<n\}\
\end{align}
\end{definition}
Notice that in this definition we have $\psi\in G\times\Sigma$ as an element of the underlying classical space while we write $|\psi\rangle\in\ell ^2(G\times\Sigma )$ to refer to the corresponding state. By using a substitution from Proposition 2.2, the generating functions may be rewritten as:
\begin{align}
f_{b|\psi}(z)=\sum _{t=1}^\infty\left(\sum _{\gamma\in F_{b|\psi}^{(t)}}A(\gamma )\right) z^t.\
\end{align}

Now that we have a pathwise representation of the generating functions, we will be able to write self-referential relations by dividing up associated paths in certain ways. We consider one such relation:
\begin{proposition}
Let $\Pi _\text{no}^BQ\leftrightarrow (G,\Sigma ,U,B)$ be an absorbing quantum walk operator and let $f_{b|\psi}(z)$ be a generating function. If $\psi =(g_0,\sigma _0)\equiv |g_0\rangle |\sigma _0\rangle$, let $\Sigma '=\{\sigma\in\Sigma :\sigma+g_0\notin B\}$. Then:
\begin{align}
f_{b|\psi}(z)=\langle b|Q|\psi\rangle z+z\sum _{\sigma\in\Sigma '}f_{b|(\sigma+g_0 )}(z).\
\end{align}
\end{proposition}
We call equation (9) in this proposition a \emph{first step transformation}. The proof of the proposition is simple and easily seen from separating the first term from the generating function in the Taylor expansion. The generating functions in the summation on the right will typically be related to the generating function on the left in some way after conducting a second transformation. This second transformation we consider is more involved:
\begin{definition}
Let $\Pi _\text{no}^BQ\leftrightarrow (G,\Sigma ,U,B)$ be an absorbing quantum walk operator and let $\psi =(g_0,\sigma _0)$. If there exists a set $X\subset G\times\Sigma$ such that, for every $n\in\N$ and $\gamma\in F_{b|\sigma}^{(n)}$, there exists a $k<n$ such that $\left( S(\theta _j(\gamma )),\varphi _j(\gamma )\right)\notin X$ for $j<k$ and $\left( S(\theta _k(\gamma )),\varphi _k(\gamma )\right)\in X$, then we say the generating function $f_{b|\psi}(z)$ is \emph{segmented} by $X$.
\end{definition}
As an illustrative simplified example, consider the set of paths $F_{0|k}^{(n)}\subset\{ -1,1\} ^n$ on the integer lattice which begin at position $k>0$ and end at position 0 at time $n$. There must exist a time $n_0\ge 1$ such that the path must intersect $k-1$ for the first time. By a simplified version of the previous definition, we say that the generating function associated with $F_{0|k}^{(n)}$ is \emph{segmented} by $k-1$. The generating functions for one dimensional quantum walks are typically segmented by a single element, but in higher dimensions the segmenting sets are larger. We now prove the \emph{segmenting transformation} in the following proposition:
\begin{proposition}
Let $\Pi _\text{no}^BQ\leftrightarrow (G,\Sigma ,U,B)$ be an absorbing quantum walk operator and let $f_{b|\psi}(z)$ be a generating function segmented by $X$. Then we have:
\begin{align}
f_{b|\psi}(z)=\sum _{\psi '\in X}f_{b|\psi '}(z)f_{\psi '|\psi}(z)\
\end{align}
\end{proposition}
{\bf Proof:} Since $f_{b|\psi}(z)$ is segmented by $X$, for every $\gamma\in F_{b|\psi}^{(n)}$ there exists a $k<n$ and an element $\psi '\in X$ such that $\theta _k(\gamma )\in F_{\psi '|\psi}^{(k)}$, and if $\theta '_k:\Sigma ^{n+1}\rightarrow\Sigma ^{k+1}$ restricts an $n$-path $\gamma =(\sigma _0,...,\sigma _n)$ to the $n-k$-path $\theta '_k(\gamma )=(\sigma _k,...,\sigma _n)$, then $\theta '_{n-k}(\gamma )\in F_{b|\psi '}^{(n-k)}$. This implies:
$$\sum _{\gamma\in F_{b|\psi}^{(n)}}A(\gamma )=\sum _{\psi '\in X}\left[\sum _{k=1}^n\left(\sum _{\gamma _1\in F_{\psi '|\psi}^{(k)}}A(\gamma _1)\right)\left(\sum _{\gamma _2\in F_{b|\psi '}^{(n-k)}}A(\gamma _2)\right)\right] .$$
The result follows from recognizing the formula of the product of two Taylor series. $\hfill\Box$

Using these two transformations in tandem allows us to write a fully self-referential relation among generating functions, upon which we can derive closed forms. For the semi-infinite walks, these closed forms may be computed explicitly. In the finite quantum walks, the generating functions satisfy recursions in the size of the lattice. These recursions typically take the form:
\begin{align}
f_{n+1}(z)=\frac{af_n(z)+b}{cf_n(z)+d}\
\end{align}
where $f_1(z)=0$. We compute a closed form expression of $f_n(z)$ in the following lemma:
\begin{lemma}
Let $\{ f_n(z)\}$ be a sequence of functions satisfying (11) and let $f_1(z)=0$. If we let $\lambda _\pm(z)=\frac{1}{2}\left[ a+d\pm\sqrt{(a+d)^2-4(ad-bc)}\right]$ and $R_n(z)=\lambda _+(z)^n-\lambda _-(z)^n$, then:
\begin{align}
f_n(z)=\frac{bR_{n-1}(z)}{R_n(z)-aR_{n-1}(z)}\
\end{align}
\end{lemma}
The proof of this lemma follows from framing this recursion as a matrix multiplication. The coefficients $\{ a,b,c,d\}$ will typically be polynomials in $z$. We will also find the following recurrence governing $R_n(z)$ useful:
\begin{align}
R_{n+2}(z)-(a+d)R_{n+1}(z)+(ad-bc)R_n(z)=0\
\end{align}

\subsection*{Hadamard Product}

In order to compute absorption probabilities from the generating functions, we must use a construction from complex analysis called the \emph{Hadamard product} \cite{titchmarsh39}.
\begin{definition}
Let $f(z)=\sum _{k=0}^\infty a_kz^k$ and $g(z)=\sum _{k=0}^\infty b_kz^k$. Then the \emph{Hadamard product} of $f$ and $g$ evaluated at $z$ is defined as follows:
\begin{align}
\left( f\odot g\right) (z)=\sum _{k=0}^\infty a_kb_kz^k\
\end{align}
\end{definition}
This Hadamard product has an integral representation which we will use extensively:
\begin{proposition}
If $\gamma$ is a contour in the $w$ plane on which $f(w)$ and $g\left(\frac{z}{w}\right)$ are analytic, then we may write the following:
\begin{align}
\left( f\odot g\right) (z)=\frac{1}{2\pi i}\int _\gamma\frac{1}{w}f(w)g\left(\frac{z}{w}\right) dw\
\end{align}
\end{proposition}
{\bf Proof:} See \cite{titchmarsh39}. $\hfill\Box$

Using this Hadamard product, one can express the absorption probabilty $P$ in terms of the generating functions. Interchanging the order of summation in equation (1) defining $P$, we may write:
\begin{align}
P=\sum _{b\in B_0}\left( f_{b|\psi}(z)\odot\overline{f_{b|\psi}(\bar{z})}\right) (1)\
\end{align}
In general, the Hadamard product is difficult to compute, but our generating functions are in a form that makes computation possible. In particular, the integrand in equation (15) of Proposition 2.5 can be divided via a partial fractions expansion. One of these fractions has all but a small number of poles contained outside the contour of integration, leading to an easy evaluation via residue theorem \cite{ahlfors53}. The other fraction has all of its poles contained inside the contour. The special form of this fraction allows us to use the following lemma:
\begin{lemma}
Let $f:\C\rightarrow\C$ be a rational function whose poles lie in a contour $\gamma$. Further, if we write $f(z)=\frac{p(z)}{q(z)}$, let $\text{deg }(q)>\text{deg }(p)+1$. Then we have:
\begin{align}
\int _\gamma f(z)dz=0\
\end{align}
\end{lemma}
{\bf Proof:} See \cite{ahlfors53}.

\section{One Dimensional Classical Random Walk}

Before addressing absorption probabilities in the quantum walk, we compute absorption probabilities for the one dimensional classical random walk to illustrate what we mean by ``classical behavior". While we opt to use language from probability theory in this section, the path counting arguments from the previous section will carry over. We need not compute any Hadamard products since the random walk is a Markov chain and not a quantum Markov chain.

We must first formally define the classical random walk:
\begin{definition}
A \emph{classical random walk} is a sequence of random variables $\{ X_j\} _{j=0}^\infty$ which satisfy $P(X_{t+1}=X_t+1)=p$ and $P(X_{t+1}=X_t-1)=q=1-p$.
\end{definition}
Notice that the quantum $n$-paths we previously defined are transition-based because the amplitude associated with a particular translation direction depends on the current state. However, in the classical random walk, probability of left/right movement is independent of previous movements. To carry over our path counting methods in this context, we must reassess the previous definition of a path:
\begin{definition}
An \emph{$n$-path} is an element $\gamma\in\{ -1,1\} ^n$ written as $\gamma =(x_1,...,x_n)$. We say that the associated \emph{displacement} of $\gamma$ is $S(\gamma )=\sum _{k=1}^nx_k$ while the associated \emph{probability} of $\gamma$ is $P(\gamma )=p^{(n+S(\gamma ))/2}q^{(n-S(\gamma ))/2}$.
\end{definition}

\subsection*{Semi-Infinite Case}

Let us consider a classical random walk for which $X_0=m>0$ and let there be an absorbing boundary at position 0. We want to calculate the absorption probability $P_\infty ^{(m)}$ that the particle is eventually absorbed at position 0. The absorption probability may be written as:
$$P_\infty ^{(m)}=\sum _{t=1}^\infty P(X_t=0|X_0=m,X_j>0\hspace{0.2cm}\forall j<t).$$
This probability has a convenient pathwise representation. If we let $\Gamma _t^{(m)}=\{\gamma\in\{ -1,1\} ^t:S(\gamma)=-m,S(\theta _k(\gamma))>-m\hspace{0.2cm}\forall k<t\}$, then we can write the following:
$$P_\infty ^{(m)}=\sum _{t=1}^\infty\left(\sum _{\gamma\in\Gamma _t^{(m)}}P(\gamma )\right) .$$

From this equation we construct a collection of generating functions:
$$f_\infty ^{(m)}(z)=\sum _{t=1}^\infty\left(\sum _{\gamma\in\Gamma _t^{(m)}}P(\gamma )\right) z^t.$$
It is clear that $P_\infty ^{(m)}=f_\infty ^{(m)}(1)$. It remains to compute a closed form for $f_\infty ^{(m)}(z)$. Since $f_\infty ^{(m)}(z)$ is segmented by $m-1$, we can use a segmenting transformation and induction to find:
$$f_\infty ^{(m)}(z)=f_\infty ^{(1)}(z)f_\infty ^{(m-1)}(z)=\left( f_\infty ^{(1)}(z)\right) ^m.$$
We have reduced this absorption problem to understanding the function $f(z)=f_\infty ^{(1)}(z)$. By a first step transformation and substituting the previous relation, we find:
$$f(z)=qz+pzf(z)^2.$$
The function $f(z)$ satisfies a quadratic equation, but there is a bit of subtlety in selecting the proper solution. To keep $f(0)$ well defined, we choose the solution:
\begin{align*}
f(z)=\begin{cases} 
      \frac{1}{2p}\left[1-\sqrt{1-4pqz^2}\right] & p\le q \\
      \frac{1}{2p}\left[1+\sqrt{1-4pqz^2}\right] & p>q
   \end{cases}.
\end{align*}
Using this closed form we are immediately granted a solution to the absorption problem:
\begin{theorem}
The absorption probabilities $P_\infty ^{(m)}$ satisfy:
\begin{align}
P_\infty ^{(m)}=\begin{cases} 
      1 & p\le q \\
      \left(\frac{q}{p}\right) ^m & p>q
   \end{cases} .\
\end{align}
\end{theorem}

\begin{figure}
\begin{center}\includegraphics[scale=0.6]{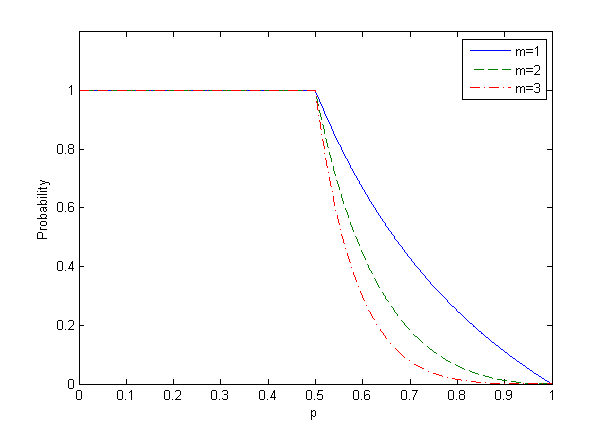}\includegraphics[scale=0.57]{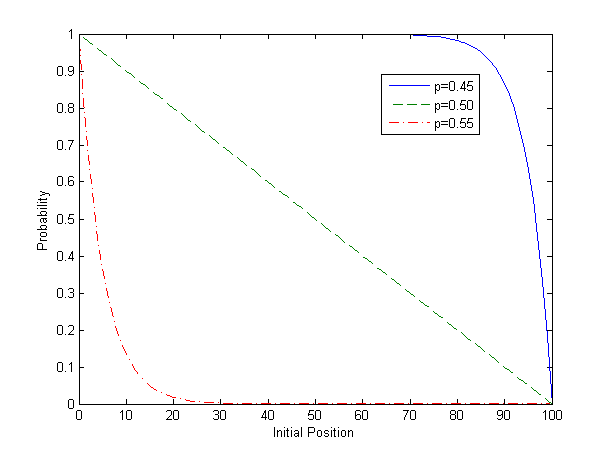}\end{center}
\caption{Left absorption probabilities for a random walk on a one dimensional lattice of size 100}
\end{figure}

\subsection*{Finite Case}

The finite case is solved in a similar way as the semi-infinite case. Let us consider a random walk for which $X_0=m$ and there are two absorbing boundaries, one at position 0 and another at position $n>m$. We wish to find the absorption probability $P_n^{(m)}$ that the particle is eventually absorbed at position 0 and not at $n$. We write this as:
$$P_n^{(m)}=\sum _{t=1}^\infty P(X_t=0|X_0=m,0<X_j<n\hspace{0.2cm}\forall j<t).$$
This probability also has a pathwise representation. Considering the path collection $\Gamma _t^{(m)}$ from the previous section, define $\Gamma ^{(m,n)}_t=\Gamma _t^{(m)}\cap\{\gamma\in\{ -1,1\} ^t:S(\theta _k(\gamma ))<n-m\hspace{0.2cm}\forall k\le t\}$. Then we can write the following:
$$P_n ^{(m)}=\sum _{t=1}^\infty\left(\sum _{\gamma\in\Gamma ^{(m,n)}_t}P(\gamma )\right) .$$

We construct the obvious class of generating functions from these absorption probabilities:
$$f_n^{(m)}(z)=\sum _{t=1}^\infty\left(\sum _{\gamma\in\Gamma ^{(m,n)}_t}P(\gamma )\right) z^t.$$
Here, it is clear that $P_n^{(m)}=f_n^{(m)}(1)$, so it remains to gather closed forms of these generating functions. The segmenting transformation is slightly different from last time; the generating function $f_n^{(m)}(z)$ is segmented by $m-1$, however via induction we have:
$$f_n^{(m)}(z)=f_{n-m+1}^{(1)}(z)f_n^{(m-1)}(z)=\prod _{k=1}^m f_{n-m+k}^{(1)}(z).$$
It thus remains to find closed forms of the generating functions for which $m=1$. By using a first step transformation and a subsequent substitution of the segmenting relation, we have:
$$f_n^{(1)}(z)=qz+pzf_n^{(1)}(z)f_{n-1}^{(1)}(z).$$
Thus we arrive at a linear fractional recurrence in $n$ governing these functions:
$$f_{n+1}^{(1)}(z)=\frac{qz}{1-pzf_n^{(1)}(z)}.$$
Here, $f_1^{(1)}(z)=0$. Using results from Lemma 2.1, we arrive at a closed form for $f_n^{(1)}(z)$:
$$f_n^{(1)}(z)=qz\frac{R_{n-1}(z)}{R_n(z)}.$$
Here, $R_n(z)=\lambda _+(z)^n-\lambda _-(z)^n$ and $\lambda _\pm (z)=\frac{1}{2}\left[ 1\pm\sqrt{1-4pqz^2}\right]$. Recognize that $\lambda _+(1)=q$ and $\lambda _-(1)=p$. If $p\ne q$, then $P_n^{(1)}=q\left(\frac{p^{n-1}-q^{n-1}}{p^n-q^n}\right)$, but if $p=q$, we find $P_n^{(1)}=1-\frac{1}{n}$. Since $P_n^{(m)}=\prod_{k=1}^m P_{n-m+k}$, a telescoping relation gives us the following:
\begin{theorem}
The absorption probabilities $P_n^{(m)}$ satisfy the following:
\begin{align}
P_n^{(m)}=\begin{cases}
	q^m\left(\frac{p^{n-m}-q^{n-m}}{p^n-q^n}\right) & p\ne q \\
      1-\frac{m}{n} & p=q
   \end{cases}\
\end{align}
\end{theorem}

\subsection*{Discussion}

Before continuing to computations involving the quantum walk, we make note of a few key characteristics these absorption probabilities share. First, notice that $P_\infty ^{(m)}=1$ if $q\le p$. If we keep the absorbing boundary at the origin and let $m$ be negative, a simple rearrangement of the previous results shows that $P_\infty ^{(-m)}=1$ for $p\le q$. Combining these results, we note that the symmetric random walk for $p=q$ is recurrent; that is, the symmetric random walk will eventually visit any point on $\Z$ with probability 1. As mentioned in the introduction, this holds for the classical random walk on $\Z ^2$ but not on $\Z ^n$ for $n\ge 3$.

Second, these absorption probabilities interchange with the limit nicely:
\begin{theorem}
For absorption probabilities of the random walk, the following relation holds:
\begin{align}
\lim _{n\rightarrow\infty} P_n^{(m)}=P_\infty ^{(m)}.\
\end{align}
\end{theorem}
While this may seem like an obvious property, it will not be satisfied for the quantum walks we discuss later. We have another result if we take the limit in a different way:
\begin{theorem}
If $0<c<1$, then:
\begin{align}
\lim _{n\rightarrow\infty} P_n^{(cn)}=\begin{cases} 
      0 & p>q \\
      1-c & p=q \\
      1 & p<q
   \end{cases}.\
\end{align}
\end{theorem}
We think of $c$ as the normalized distance between the left and right absorbing boundaries. We include this theorem because these absorption probabilities are stable in the limit taken on a ray $cn$. In particular, there is a dichotomy between the cases guaranteed absorption to the left ($p<q$) and guaranteed absorption to the right ($p>q$) where the choice of ray has no impact on the probability, and the symmetric case in which the choice of ray actually does affect the absorption probability.

Lastly, we write a few recurrence relations among the absorption probabilities. The proof of these relationships is straightforward:
\begin{theorem}
Let $P_n^{(m)}$ be an absorption probability of the random walk. Then the following formulas hold:
\begin{align}
P_{n+1}^{(1)}=\frac{q}{1-pP_n^{(1)}};\
\end{align}
\begin{align}
P_n^{(m)}=\prod _{k=1}^m P_{n-m+k}^{(1)};\
\end{align}
\begin{align}
P_n^{(m+2)}-\frac{1}{p}P_n^{(m+1)}+\frac{q}{p}P_n^{(m)}=0.\
\end{align}
\end{theorem}
{\bf Proof:} The first two formulas hold by letting $z=1$ in two of the generating function relations above. The third may be derived by noting that $R_{n+2}-R_{n+1}+pqR_n=0$ and multiplying this relation by $q^{m+1}$. $\hfill\Box$

\section{$(\Z ,C_1,U)$ Quantum Walk}

We now compute absorption probabilities for the $(\Z ,C_1,U)$ quantum walk where $C_1=\{ -1,1\}$ and $U$ is an arbitrary $2\times 2$ unitary matrix. It suffices to consider matrices of the form $U=\begin{bmatrix} a & b \\ -\bar{b} & \bar{a}\end{bmatrix}$, as phase constants will not affect the absorption probabilities. As mentioned previously, the case of $U=H=\frac{1}{\sqrt{2}}\begin{bmatrix} 1 & 1 \\ 1 & -1\end{bmatrix}$ where $H$ is the $2\times 2$ Hadamard matrix has been studied in a few papers \cite{ambainis01} \cite{bach04} \cite{bach09}.

\subsection*{Semi-Infinite Case}

We first consider the absorption probabilities $P_\infty ^{(m)}(\alpha ,\beta )$ corresponding to an initial condition $|m\rangle (\alpha |R\rangle +\beta |L\rangle )$ and the quantum walk operator $\Pi _\text{no}^0Q\leftrightarrow (\Z ,C_1,U,\{ 0\} )$. If we place an absorbing boundary at $|0\rangle$, we need only consider absorption at $|0\rangle |L\rangle$ as there will never be amplitude at $|0\rangle |R\rangle$ in this setting. We can write out the absorption probabilities corresponding to arbitrary initial condition as follows:
\begin{align}
P_\infty ^{(m)}(\alpha ,\beta ) &= \sum _{t=1}^\infty |\alpha |^2|\langle 0,L|Q(\Pi _\text{no}^0Q)^{t-1}|m,R\rangle |^2+\sum _{t=1}^\infty |\beta |^2|\langle 0,L|Q(\Pi _\text{no}^0Q)^{t-1}|m,L\rangle |^2\nonumber\ \\
&+\sum _{t=1}^\infty 2\text{Re }\left[\alpha\bar{\beta}\langle 0,L|Q(\Pi _\text{no}^0Q)^{t-1}|m,R\rangle\overline{\langle 0,L|Q(\Pi _\text{no}^0Q)^{t-1}|m,L\rangle}\right] .\
\end{align}
Recognize the first two quantities in this sum as $|\alpha |^2P_\infty ^{(m)}(1,0)+|\beta |^2P_\infty ^{(m)}(0,1)$. If we define
$$H_\infty ^{(m)}=\sum _{t=1}^\infty\langle 0,L|Q(\Pi _\text{no}^0Q)^{t-1}|m,R\rangle\overline{\langle 0,L|Q(\Pi _\text{no}^0Q)^{t-1}|m,L\rangle} ,$$
then we may rewrite the general absorption probability as:
\begin{align}
P_\infty ^{(m)}(\alpha ,\beta )=|\alpha |^2P_\infty ^{(m)}(1,0)+|\beta |^2P_\infty ^{(m)}(0,1)+2\text{Re }\left[\alpha\bar{\beta}H_\infty ^{(m)}\right] .\
\end{align}
While the calculation of generating functions is manageable for general $m$, computation of the Hadamard product for these cases is unwieldy. We thus illustrate the calculation of $P_\infty ^{(1)}(1,0)$. This is readily extended to the general absorption probability $P_\infty ^{(1)}(\alpha ,\beta )$ by equation (39) below.

\subsubsection*{Generating Functions}

We first define the following two generating functions:
$$r_\infty ^{(m)}(z)=\sum _{t=1}^\infty \langle 0,L|Q(\Pi _\text{no}^0 Q)^{t-1}|m,R\rangle z^t,\hspace{1cm}l_\infty ^{(m)}(z)=\sum _{t=1}^\infty \langle 0,L|Q(\Pi _\text{no}^0 Q)^{t-1}|m,L\rangle z^t.$$
We use this subsection to prove the following proposition:
\begin{proposition}
The generating functions $r_\infty ^{(m)}(z)$ and $l_\infty ^{(m)}(z)$ have the following closed forms:
\begin{align}
r_\infty ^{(m)}(z)&= \frac{a}{b}\left(\frac{1}{2az}\right) ^m\left( 1-z^2-\sqrt{z^4+2(|b|^2-|a|^2)z^2+1}\right)\nonumber \\
	&\times\left( 1+z^2-\sqrt{z^4+2(|b|^2-|a|^2)z^2+1}\right) ^{m-1}
\end{align}
\begin{align}
l_\infty ^{(m)}(z)=\left(\frac{1}{2az}\right) ^m\left( 1+z^2-\sqrt{z^4+2(|b|^2-|a|^2)z^2+1}\right) ^m\
\end{align}
\end{proposition}
{\bf Proof:} Both of these generating functions are segmented by $|m-1\rangle |L\rangle$. Using a segmenting transformation and induction, the following relationships hold:
$$r_\infty ^{(m)}(z)=r_\infty ^{(1)}(z)\left( l_\infty ^{(1)}(z)\right) ^{m-1},\hspace{1cm}l_\infty ^{(m)}(z)=\left( l_\infty ^{(1)}(z)\right) ^m.$$
It now remains to handle the $m=1$ generating functions. Using a first step transformation and substituting this segmenting relation, we arrive at the system:
$$r_\infty ^{(1)}(z)=-\bar{b}z+azr_\infty ^{(1)}(z)l_\infty ^{(1)}(z),\hspace{1cm}l_\infty ^{(1)}(z)=\bar{a}z+bzr_\infty ^{(1)}(z)l_\infty ^{(1)}(z).$$
This system can be solved simply and results in the following solutions:
$$r_\infty ^{(1)}(z)=\frac{1}{2bz}\left[ 1-z^2-\sqrt{z^4+2(|b|^2-|a|^2)z^2+1}\right] .$$
\begin{align}
l_\infty ^{(1)}(z)=\frac{1}{2az}\left[ 1+z^2-\sqrt{z^4+2(|b|^2-|a|^2)z^2+1}\right] .\
\end{align}
We choose the negative square root so that these generating functions have a convergent Taylor series about $z=0$. The result follows upon substitution. $\hfill\Box$

\subsubsection*{Hadamard Product}

The absorption probabilities may be written in terms of the generating functions as follows:
$$P_\infty ^{(m)}(1,0)=\left( r_\infty ^{(m)}(z)\odot\overline{r_\infty ^{(m)}(\bar{z})}\right) (1),\hspace{0.5cm}P_\infty ^{(m)}(0,1)=\left( l_\infty ^{(m)}(z)\odot\overline{l_\infty ^{(m)}(\bar{z})}\right) (1) .$$
\begin{align}
H_\infty ^{(m)}=\left( r_\infty ^{(m)}(z)\odot\overline{l_\infty ^{(m)}(\bar{z})}\right) (1) .\
\end{align}

Presently, we are only able to state the semi-infinite absorption probabilities for the $m=1$ case. For the $U=H$ case, Ambainis et.\ al.\ used a combinatorial appeal to the Catalan numbers to prove that $P_\infty ^{(1)}(1,0)=\frac{2}{\pi}$. However, this approach does not generalize well to the arbitrary unitary matrix case and we opt to use the integral expression of the Hadamard product in our proof.
\begin{theorem}
The following formula holds:
\begin{align}
P_\infty ^{(1)}(1,0)=\frac{2}{\pi |b|^2}\left( (|b|^2-|a|^2)\cos ^{-1}(|a|)+|a||b|\right)\
\end{align}
Here, $\cos ^{-1}:[-1,1]\rightarrow\left[ 0,\pi\right]$. If we write $|a|=\cos\phi$ where $\phi\in [0,\frac{\pi}{2}]$, then we also have:
\begin{align}
P_\infty ^{(1)}(1,0)=\frac{\sin{2\phi}+2\phi\cos{2\phi}}{\pi\sin ^2\phi}\
\end{align}
\end{theorem}
{\bf Proof:} From Ambainis et.\ al.\ we are able to alter the integral representation of the Hadamard product slightly:
$$P_\infty ^{(1)}=\frac{1}{2\pi}\int _0^{2\pi}|r_\infty ^{(1)}(e^{i\theta})|^2d\theta .$$
For simplicity, let $\alpha =|b|^2-|a|^2$. By expanding $|r_\infty ^{(1)}(e^{i\theta})|^2$, we have:
\begin{align*}
P_\infty ^{(1)} &= \frac{1}{8\pi |b|^2}\left(\int_0^{2\pi}|1-e^{2i\theta}|^2d\theta +\int _0^{2\pi}|e^{4i\theta}+2\alpha e^{2i\theta}+1|d\theta -2\text{Re }\left[\int_0^{2\pi}(1-e^{-2i\theta})\sqrt{e^{4i\theta}+2\alpha e^{2i\theta}+1}d\theta\right]\right) \\
	&=\frac{1}{8\pi |b|^2}(I_1+I_2-2\text{Re }(I_3))
\end{align*}
Through straightforward integration, it is easy to prove that $I_1=4\pi$. Notice that $f(\theta ):=e^{4i\theta}+2\alpha e^{2i\theta}+1=2(\alpha +\cos{2\theta})e^{2i\theta}$. Letting $\phi\in (0,\frac{\pi}{2})$ satisfy $\cos{2\phi}=-\alpha$, we can solve for $I_2$:
\begin{align*}
I_2 &= 2\int _0^{2\pi} |\alpha +\cos{2\theta}| d\theta \\
	&= 4\left(\int _{-\phi}^\phi \alpha+\cos{2\theta}d\theta-\int_\phi ^{\pi -\phi}\alpha +\cos{2\theta}d\theta\right) \\
	&= 4(4\alpha\phi +4|a||b|-\alpha\pi)
\end{align*}
Here, we use the identity $\sin{2\phi}=\sqrt{1-\alpha ^2}=2|a||b|$. For the final integral $I_3$, notice that $f(\theta )$ traces a rose curve \cite{cundy61} variant in the complex plane which only intersects the branch cut $(-\infty ,0]$ at the branch point, thus making this integral well defined. Let us choose a different branch of the square root to operate on real numbers (denoted by $\sqrt[\R]{\cdot}$) such that square root of a positive real number to have positive real part and the square root of a negative real number has positive imaginary part. We can rewrite $\sqrt{f(\theta )}$ (in the former sense) on the interval $\theta\in [0,2\pi ]$ as:
$$\sqrt{f(\theta )}=\sqrt[\R]{2(\alpha +\cos{2\theta})}e^{i\theta}\left(\chi _{[0,\phi ]}-\chi _{[\phi ,\pi +\phi ]}+\chi _{[\pi +\phi ,2\pi ]}\right)$$
Notice that the integrand of $I_3$ is purely imaginary when $\alpha +\cos{2\theta}>0$ and purely real otherwise. Thus, we can use the indefinite integral
$$\int\sqrt{\alpha +\cos{2\theta}}\sin{\theta}d\theta =-\frac{1}{4}\left( 2\cos\theta\sqrt{\alpha +\cos{2\theta}}-\sqrt{2}(\alpha -1)\log\left(\sqrt{\alpha +\cos{2\theta}}+\sqrt{2}\sin\theta\right)\right)$$
to find that $2\text{Re }(I_3)=8\pi |a|^2$. The result follows. $\hfill\Box$

\subsection*{Finite Case}

Let $\Pi _\text{no}^{n}\Pi _\text{no}^{0}Q\leftrightarrow (\Z ,C_1,U,\{0 ,n\})$ be an absorbing quantum walk operator corresponding to the quantum walk on the finite one dimensional lattice. We are intereseted in the absorption probabilities $P_n^{(m)}(\alpha ,\beta )$ or the probability that a quantum walk particle initialized in $|m\rangle (\alpha |R\rangle +\beta |L\rangle )$ is eventually absorbed at $|0\rangle$ before being absorbed at $|n\rangle$. We can divide the general absorption probability as we had in the previous section:
\begin{align}
P_n^{(m)}(\alpha ,\beta )=|\alpha |^2P_n^{(m)}(1,0)+|\beta |^2P_n^{(m)}(0,1)+2\text{Re }\left[\alpha\bar{\beta}H_n^{(m)}\right].\
\end{align}
Again, we will be focusing our attention on computing $P_n^{(m)}(1,0)$, as the other two quantities are calculated similarly.

\subsubsection*{Generating Functions}

Let us define the generating functions:
$$r_n ^{(m)}(z)=\sum _{t=1}^\infty \langle 0,L|Q(\Pi _\text{no}^n\Pi _\text{no}^0 Q)^{t-1}|m,R\rangle z^t,\hspace{1cm}l_n ^{(m)}(z)=\sum _{t=1}^\infty \langle 0,L|Q(\Pi _\text{no}^n\Pi _\text{no}^0 Q)^{t-1}|m,L\rangle z^t .$$
This section is devoted to computing the following closed form of these generating functions:
\begin{proposition}
We have:
\begin{align}
r_n^{(m)}(z)=\frac{-\bar{b}\bar{a}^{m-1}z^mR_{n-m}(z)}{R_n(z)-z^2R_{n-1}(z)},\hspace{1cm}l_n^{(m)}(z)=\bar{a}^mz^m\left(\frac{R_{n-m}(z)-z^2R_{n-m-1}(z)}{R_n(z)-z^2R_{n-1}(z)}\right) .\
\end{align}
\end{proposition}
{\bf Proof:} Both of these generating functions are segmented by $|m-1\rangle |L\rangle$. Using a combination of the segmenting transformation and induction, we may write the generating functions using $m=1$ generating functions:
$$r_n^{(m)}(z)=r_{n-m+1}^{(1)}(z)\prod _{k=2}^ml_{n-m+k}^{(1)}(z),\hspace{1cm}l_n^{(m)}(z)=\prod _{k=1}^ml_{n-m+k}^{(1)}(z).$$
Using a first step transformation along with this formula, we arrive at the following system:
$$r_n^{(1)}(z)=-\bar{b}z+azr_{n-1}^{(1)}(z)l_n^{(1)}(z),\hspace{1cm}l_n^{(1)}(z)=\bar{a}z+bzr_{n-1}^{(1)}(z)l_n^{(1)}(z) .$$
These two formulas can be combined to construct a recursion for $r_n^{(1)}(z)$:
$$r_{n+1}^{(1)}(z)=\frac{z^2r_n^{(1)}(z)-\bar{b}z}{bzr_n^{(1)}(z)-1}.$$
By using lemma 2.1, we can solve for the following closed form:
$$r_n^{(1)}(z)=\frac{-\bar{b}zR_{n-1}(z)}{R_n(z)-z^2R_{n-1}(z)} .$$
Here, $R_n(z)=\lambda _+(z)^n-\lambda _-(z)^n$ and $\lambda _\pm (z)=\frac{1}{2}\left[ (1+z^2)\pm\sqrt{z^4+2(|b|^2-|a|^2)z^2+1}\right]$. Meanwhile, by recognizing $l_n^{(1)}(z)=\frac{b}{a}r_n^{(1)}(z)+\frac{z}{a}$ and using the recursion relation $R_{n+1}(z)-z^2R_n(z)=R_n(z)-|a|^2z^2R_{n-1}(z)$, we can calculate:
$$l_n^{(1)}(z)=\bar{a}z\left(\frac{R_{n-1}(z)-z^2R_{n-2}(z)}{R_n(z)-z^2R_{n-1}(z)}\right) .$$
Applying induction and the segmenting relation derives the result. $\hfill\Box$

\subsubsection*{Hadamard Product}

We will detail the computation of $P_n^{(m)}(1,0)=\left( r_n^{(m)}(z)\odot\overline{r_n^{(m)}(\bar{z})}\right) (1)$. We wish to use the integral representation of this Hadamard product as given by proposition 2.5, so we must provide a region for which the generating functions are analytic. This requires a lemma:
\begin{lemma}
Let $f(w,z)=\frac{z^2w-\bar{b}z}{-bzw+1}$ where $|b|<1$. Then $|w|,|z|\le 1\Rightarrow |f(w,z)|\le 1$.
\end{lemma}
{\bf Proof:} Let us rewrite $f(w,z)$ as:
$$f(w,z)=\frac{w-\frac{\bar{b}}{z}}{-bzw+1} .$$
First consider the function $g_z(w)=f(w,z)$ where $z$ is fixed. If $|z|=1$, then it follows that $g_z(w)$ is an automorphism of the unit disk. Next consider the function $h_w(z)=f(w,z)$ where $w$ is fixed. For $|w|\le 1$, $h_w(z)$ is analytic in the unit disk since there is a single pole located at $z=\frac{1}{bw}$. The result now follows from the maximum modulus principle. $\hfill\Box$

This lemma allows us to prove the following proposition:
\begin{proposition}
For every $m,n\in\N$, there exists an $\epsilon >0$ such that $r_n^{(m)}(z)$ and $l_n^{(m)}(z)$ are analytic in the disk $|z|<1+\epsilon$.
\end{proposition}
{\bf Proof:} Via induction, the previous lemma shows that $|z|\le 1\Rightarrow |r_n^{(1)}(z)|\le 1$ for all $n\in\N$, and a continuity argument tells us that for every $n\in\N$ there exists an $\epsilon >0$ such that $r_n^{(1)}(z)$ is analytic in the disk $|z|<1+\epsilon$. Since $l_n^{(1)}(z)=\frac{b}{a}r_n^{(1)}(z)+\frac{z}{a}$, this conclusion must also hold for $l_n^{(1)}(z)$. Since $r_n^{(m)}(z)$ and $l_n^{(m)}(z)$ are products of these $m=1$ generating functions, the result follows. $\hfill\Box$

Before moving onto the computation of the Hadamard product, we prove two lemmas which will aid in this endeavor:
\begin{lemma}
The following formulas hold:
\begin{align}
\overline{r_n^{(m)}\left(\frac{1}{\bar{z}}\right)}=\frac{-ba^{m-1}z^mR_{n-m}(z)}{R_n(z)-R_{n-1}(z)},\hspace{1cm}\overline{l_n^{(m)}\left(\frac{1}{\bar{z}}\right)}=a^mz^m\frac{R_{n-m}(z)-R_{n-m-1}(z)}{R_n(z)-R_{n-1}(z)}\
\end{align}
\end{lemma}
{\bf Proof:} The result follows from noting that $\overline{\lambda _\pm\left(\frac{1}{\bar{z}}\right)}=\frac{1}{z^2}\lambda _\mp (z)$ and $\overline{R_n\left(\frac{1}{\bar{z}}\right)}=\left(\frac{1}{z^2}\right) ^nR_n(z)$. $\hfill\Box$
\begin{lemma}
The following formula holds:
$$\frac{R_{n-m}(z)}{(R_n(z)-z^2R_{n-1}(z))(R_n(z)-R_{n-1}(z))}$$
\begin{align}
=\frac{1}{|a|^{2m-2}z^{2m-2}(1-z^2)R_1(z)}\left[\frac{R_m(z)-R_{m-1}(z)}{R_n(z)-R_{n-1}(z)}-\frac{R_m(z)-z^2R_{m-1}(z)}{R_n(z)-z^2R_{n-1}(z)}\right] .\
\end{align}
\end{lemma}
{\bf Proof:} Let us assume the following relation holds:
$$\frac{R_{n-m}(z)}{(R_n(z)-z^2R_{n-1}(z))(R_n(z)-R_{n-1}(z))}=\frac{A}{R_n(z)-z^2R_{n-1}(z)}+\frac{B}{R_n(z)-R_{n-1}(z)} .$$
We can write $x_1R_n(z)+y_1R_{n-1}(z)=x_2R_{n-1}(z)+y_2R_{n-2}(z)$ where the coefficients are related to each other by the matrix equation:
$$\begin{bmatrix} x_2 \\ y_2\end{bmatrix}=\begin{bmatrix} z^2+1 & 1 \\ -|a|^2z^2 & 0\end{bmatrix}\begin{bmatrix} x_1 \\ y_1\end{bmatrix} .$$
Using an eigenvalue expansion, we find that:
$$R_{n-m}(z)=(A+B)R_n(z)-(A+z^2B)R_{n-1}(z)$$
$$= \frac{1}{R_1(z)}\left[ A(R_m(z)-R_{m-1}(z))+B(R_m(z)-z^2R_{m-1}(z))\right] R_{n-m+1}(z)\nonumber$$
$$- \frac{|a|^2z^2}{R_1(z)}\left[ A(R_{m-1}(z)-R_{m-2}(z))+B(R_{m-1}(z)-z^2R_{m-2}(z))\right] R_{n-m}(z) .$$
By solving this system and using the equation $R_m(z)^2-R_{m+1}(z)R_{m-1}(z)=|a|^{2m}z^{2m}R_1(z)^2$, the result follows. $\hfill\Box$

We are now ready to prove the main theorem of this section:
\begin{theorem}
Let $R_n=R_n(1)=(1+|b|)^n-(1-|b|)^n$ and $B_n=(1+|b|)^n+(1-|b|)^n$. The following formulas hold:
$$P_n^{(m)}(1,0)=\frac{|b|}{2}\left(\frac{R_{n-m}B_{m-1}}{B_{n-1}}\right) ,\hspace{1cm}P_n^{(m)}(0,1)=\frac{1}{2}\left(\frac{R_{n-m-1}R_m}{B_{n-1}}+|a|^2\frac{B_{n-m-1}B_{m-1}}{B_{n-1}}\right)$$
\begin{align}
H_n^{(m)}=-a\bar{b}\left(\frac{B_{n-m-1}B_{n-1}}{B_{n-1}}\right) .\
\end{align}
\end{theorem}
{\bf Proof:} We may write the absorption probability as $P_n^{(m)}(1,0)=\left( r_n^{(m)}(z)\odot\overline{r_n^{(m)}(\bar{z})}\right) (1)$. By combining proposition 2.5 and proposition 4.3, we find that there exists an $\epsilon >0$ such that:
$$P_n^{(m)}(1,0)=\frac{1}{2\pi i}\int _{|z|=1+\epsilon}\frac{1}{z}r_n^{(m)}(z)\overline{r_n^{(m)}\left(\frac{1}{\bar{z}}\right)}dz.$$
Substituting the above lemmas into this equation, we may write:
$$P_n^{(m)}(1,0)=\frac{1}{2\pi i}\int _{|z|=1+\epsilon}\frac{|b|^2|a|^{2m-2}z^{2m-1}R_{n-m}(z)^2}{(R_n(z)-z^2R_{n-1}(z))(R_n(z)-R_{n-1}(z))}dz.$$
Using the partial fractions expansion from lemma 4.3, we have:
$$P_n^{(m)}(1,0)=\frac{1}{2\pi i}\int _{|z|=1+\epsilon}\frac{|b|^2zR_{n-m}(z)}{(1-z^2)R_1(z)}\left[\frac{R_m(z)-R_{m-1}(z)}{R_n(z)-R_{n-1}(z)}-\frac{R_m(z)-z^2R_{m-1}(z)}{R_n(z)-z^2R_{n-1}(z)}\right] dz.$$
By Lemma 2.2, the partial fraction on the right may be eliminated leading to the simplified expression:
$$P_n^{(m)}(1,0)=\frac{1}{2\pi i}\int _{|z|=1+\epsilon}\frac{|b|^2zR_{n-m}(z)}{(z^2-1)R_1(z)}\left[\frac{R_m(z)-z^2R_{m-1}(z)}{R_n(z)-z^2R_{n-1}(z)}\right] dz.$$
The result follows from an application of the residue theorem as well as recognizing that $R_n-R_{n-1}=|b|B_{n-1}$. Calculation of $P_n^{(m)}(0,1)$ and $H_n^{(m)}$ is similar. $\hfill\Box$

\begin{figure}
\begin{center}\includegraphics[scale=0.9]{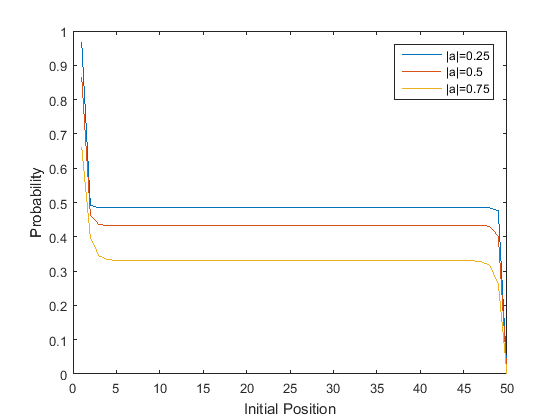}\end{center}
\caption{Plot of absorption probabilities $P_n^{(m)}(1,0)$ for $n=50$}
\end{figure}

\subsection*{Discussion}

We note several differences between this quantum walk and the classical random walk. First, note that $P_\infty ^{(1)}(\alpha ,\beta )<1$ unless $|-\bar{b}\alpha +\bar{a}\beta |=1$ (i.e. all probability is absorbed in the first step). This stands in contrast to the classical random walk which was recurrent in the symmetric case.

Second, the finite absorption probabilities do not limit to the semi-infinite absorption probabilities and this limit fails in a surprising way.
\begin{theorem}
The following inequality is satisfied:
\begin{align}
P_\infty ^{(1)}(\alpha ,\beta )\le P_n^{(1)}(\alpha ,\beta ).\
\end{align}
\end{theorem}
{\bf Proof:} First note that a walk initialized in $|1\rangle (\alpha |R\rangle +\beta |L\rangle )$ will be absorbed at time $t=1$ with probability $|-\bar{b}\alpha +\bar{a}\beta |^2$ and will have a subsequent state of $(a\alpha +b\beta )|2\rangle |R\rangle$, thus:
$$P_n^{(1)}(\alpha ,\beta )=|-\bar{b}\alpha +\bar{a}\beta |^2+|a\alpha +b\beta |^2P_n^{(2)}(1,0)$$
By substituting $\alpha =1$ and $\beta =0$, we find that $P_n^{(1)}(\alpha ,\beta )$ and $P_n^{(1)}(1,0)$ are both linear combinations of $P_n^{(2)}(1,0)$. This allows us to solve for a linear relation between the two absorption probabilities:
\begin{align}
P_n^{(1)}(\alpha ,\beta )=1-\frac{|a\alpha +b\beta |^2}{|a|^2}(1-P_n^{(1)}(1,0)) .\
\end{align}
Thus, it suffices to prove this inequality for $P_n^{(1)}(1,0)$. Note that this argument still holds for $n=\infty$ as no reference was made to the absorbing boundary. By using a trigonometric representation of the governing unitary matrix, proving the theorem becomes equivalent to proving
$$f(\theta )=\pi\sin^3{\theta}+2\theta\cos{2\theta}-\sin{2\theta}\ge 0$$
where $\theta\in [0,\frac{\pi}{2}]$. The only values of $\theta$ in this interval such that $f'(\theta )=0$ are $\theta\in\{ 0,\theta _0,\frac{\pi}{2}\}$ where $\theta _0=\frac{3\pi}{8}\sin{\theta _0}$. It can be shown that $\frac{\pi}{4}<\theta _0$ and the result follows from noting that $f'(\frac{\pi}{4})>0$. $\hfill\Box$
\begin{figure}
\begin{center}\includegraphics[scale=0.61]{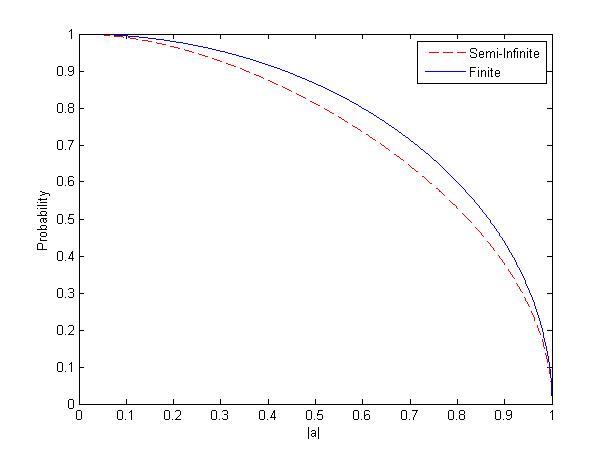}\includegraphics[scale=0.61]{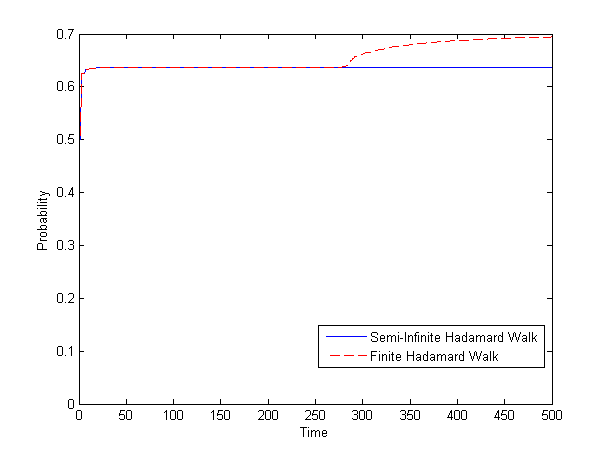}\end{center}
\caption{(\emph{Left}) Comparison of absorption probabilities $P_\infty ^{(1)}(1,0)$ and $\lim _{n\rightarrow\infty}P_n^{(m)}(1,0)$ (\emph{Right}) Cumulative distribution function of the absorption probabilities $P_\infty ^{(1)}(1,0)$ and $P_n^{(1)}(1,0)$ for $|a|=\frac{1}{\sqrt{2}}$ and $n=100$}
\end{figure}

This is perhaps the most surprising result from the initial foray into absorption probabilities by Ambainis et.\ al.\ and here we have extended it to arbitrary internal state and arbitrary governing matrix. Not only do the finite probabilities fail to limit to the semi-infinite case, but the presence of an absorbing boundary sufficiently far to the right from the initial position will actually \emph{increase} the probability of eventually being absorbed to the left. This apparent paradox can be reconciled by considering the quantum walk to be a kind of discretization of the transport equation and that these absorbing boundaries do not necessarily absorb all information but reflect some as well. Figure 3 illustrates this concept. We present a conjecture to extend this theorem:
\begin{conjecture}
Theorem 4.3 is satisfied for arbitrary initial position $m$.
\end{conjecture}
We specify these limits in the finite case as we have in the classical random walk:
\begin{theorem}
Let $0<c<1$. The following limiting probabilities hold:
\begin{align}
\lim _{n\rightarrow\infty} P_n^{(m)}(1,0)=\frac{|b|}{2}\left[1+\left(\frac{1-|b|}{1+|b|}\right) ^{m-1}\right]\
\end{align}
\begin{align}
\lim _{n\rightarrow\infty} P_n^{(cn)}(1,0)=\frac{|b|}{2}.\
\end{align}
\end{theorem}

Another result which we dsiplay proves that a quantum walk particle in the finite absorption setting is guaranteed to eventually be absorbed by one of the boundaries:
\begin{theorem}
Let $P_n^{(m)}(\alpha ,\beta ;a,b)$ be the probability that a $(\Z ,C_1,U,\{ 0,n\} )$ absorbing quantum walk particle initialized at $|m\rangle\left(\alpha |R\rangle +\beta |L\rangle\right)$ is eventually absorbed at $|0\rangle |L\rangle$ before being absorbed at $|n\rangle |R\rangle$, where $U=\begin{bmatrix} a & b \\ -\bar{b} & \bar{a}\end{bmatrix}$. Then the following formula holds:
\begin{align}
P_n^{(m)}(\alpha ,\beta ;a,b)+P_n^{(n-m)}(\beta ,\alpha ;\bar{a},-\bar{b})=1.\
\end{align}
\end{theorem}
{\bf Proof:} Consider the following quantities:
$$W_n^{(m)}=\frac{B_{n-m-1}B_{m-1}}{B_{n-1}},\hspace{0.3cm}X_n^{(m)}=\frac{R_{n-m-1}R_{m-1}}{B_{n-1}},\hspace{0.3cm}Y_n^{(m)}=\frac{R_{n-m-1}B_{m-1}}{B_{n-1}},\hspace{0.3cm}Z_n^{(m)}=\frac{B_{n-m-1}R_{m-1}}{B_{n-1}}$$
By using the index reducing formulas (i.e. $R_n=R_{n-1}+|b|B_{n-1}$ and $B_n=B_{n-1}+|b|R_{n-1}$) on the quantities in equation (37), we have the following representations:
$$P_n^{(m)}(1,0)=\frac{|b|}{2}Y_n^{(m)}+\frac{|b|^2}{2}W_n^{(m)},\hspace{0.7cm}P_n^{(m)}(0,1)=\frac{1}{2}X_n^{(m)}+\frac{|b|}{2}Y_n^{(m)}+\frac{|a|^2}{2}W_n^{(m)},\hspace{0.7cm}H_n^{(m)}=-a\bar{b}W_n^{(m)}$$
Notice that $W_n^{(n-m)}=W_n^{(m)}$ and $X_n^{(n-m)}=X_n^{(m)}$, while $Y_n^{(n-m)}=Z_n^{(m)}$. Also, since $W$, $X$, $Y$, and $Z$ are dependent on $|a|$ and $|b|$, they are invariant to multiplication by phase and complex conjugation of either $a$ and $b$. These observations allow us to plug equation (28) into equation (42) and collect terms:
$$P_n^{(m)}(\alpha ,\beta ;a,b)+P_n^{(n-m)}(\beta ,\alpha ;\bar{a},-\bar{b})=\frac{1}{2}(W_n^{(m)}+X_n^{(m)})+\frac{|b|}{2}(Y_n^{(m)}+Z_n^{(m)})$$
By using the identities $R_nB_m+B_nR_m=2R_{n+m}$ and $R_nR_m+B_nB_m=2B_{n+m}$, we further reduce this to:
$$P_n^{(m)}(\alpha ,\beta ;a,b)+P_n^{(n-m)}(\beta ,\alpha ;\bar{a},-\bar{b})=\frac{B_{n-2}+|b|R_{n-2}}{B_{n-1}}$$
A reverse application of the index reducing formulas completes the proof. $\hfill\Box$

It is clear in the context of the theorem that $P_n^{(m)}(\alpha ,\beta ;a,b)$ is the probabilty of left absorption, and that $P_n^{(n-m)}(\beta ,\alpha ;\bar{a},-\bar{b})$ is the probability of right absorption. 

We conclude this section by proving two recurrences of these absorption probabilities:
\begin{theorem}
The following recurrences hold for finite absorption probabilities:
\begin{align}
P_{n+1}^{(1)}(1,0)=\frac{|b|+P_n^{(1)}(1,0)}{1+P_n^{(1)}(1,0)}\
\end{align}
\begin{align}
P_n^{(m+3)}(\alpha ,\beta )-\left(\frac{4}{|a|^2}-1\right) P_n^{(m+2)}(\alpha ,\beta )+\left(\frac{4}{|a|^2}-1\right) P_n^{(m+1)}(\alpha ,\beta )-P_n^{(m)}(\alpha, \beta )=0 .\
\end{align}
\end{theorem}
{\bf Proof:} The first equation can easily be proven true by recognizing that $\begin{bmatrix} R_{n+1} \\ B_{n+1}\end{bmatrix}=\begin{bmatrix} 1 & |b| \\ |b| & 1\end{bmatrix}\begin{bmatrix} R_n \\ B_n\end{bmatrix}$. For the second equation, notice that $P_n^{(m)}$ is a linear combination of terms of the form $\{ R_{n-m}B_m,R_{n-m}R_m,B_{n-m}B_m,B_{n-m}R_m\}$. If we let
$$V_m=[R_{n-m}B_m,R_{n-m}R_m,B_{n-m}B_m,B_{n-m}R_m]'$$
then the following equation holds:
$$V_{m+1}=\frac{1}{|a|^2}\begin{bmatrix} 1 & |b| & -|b| & -|b|^2 \\ |b| & 1 & -|b|^2 & -|b| \\ -|b| & -|b|^2 & 1 & |b| \\ -|b|^2 & -|b| & |b| & 1\end{bmatrix}V_m.$$
The characteristic polynomial of this matrix is:
$$p(z)=(z-1)^2\left( z^2+2\left( 1-\frac{2}{|a|^2}\right) z+1\right) .$$
The conclusion follows from noting that the minimal polynomial retains only one of the factors of $(z-1)$. $\hfill\Box$

\section{$(\Z ,\tilde{C}_1,G_3)$ Quantum Walk}

We repeat the analysis in the previous section for the $(\Z ,\tilde{C}_1,G_3)$ quantum walk where $\tilde{C}_1=\{ -1,0,1\}$ and $G_3=\frac{1}{3}\begin{bmatrix} -1 & 2 & 2 \\ 2 & -1 & 2 \\ 2 & 2 & -1\end{bmatrix}$ is the $3\times 3$ Grover matrix. This is also known as the three-state Grover walk and its unbounded behavior has been studied in several papers \cite{inui05}\cite{falkner14}. The three-state Grover walk exhibits a property known as \emph{localization} in which the time averaged probability of observing the particle at its initial condition limits to a nonzero value. This is caused by the degeneration of eigenvalues in the Grover matrix, and it has been shown that localization exists for other quantum walks governed by Grover matrices \cite{inui04}.

These absorption probabilities have partially been computed in Wang et.\ al.\ \cite{wang16} but we repeat the analysis here in a more concise framework. We refer back to the previous section for several of the proofs.

\subsection*{Semi-Infinite Case}

In this section we consider the absorption probabilities $P_\infty ^{(m)}(\alpha ,\beta ,\gamma )$ or the probability that a particle initialized in $|m\rangle (\alpha |R\rangle +\beta |S\rangle +\gamma |L\rangle )$ is eventually absorbed in the $\Pi _\text{no}^0 Q\leftrightarrow (\Z ,\tilde{C}_1,G_3,\{ 0\} )$ absorbing quantum walk. Here, recognize that $T:|n\rangle |S\rangle\rightarrow |n\rangle |S\rangle$ for the translation operator associated with $Q$. As with the semi-infinite two state quantum walk, we are able to compute generating functions for general $m$ but restrict the Hadamard product to $m=1$.

While the generalized absorption probability $P_\infty ^{(m)}(\alpha ,\beta )$ of the previous section depended only on the three quantities $P_\infty ^{(m)}(1,0)$, $P_\infty ^{(m)}(0,1)$, and $H_n^{(m)}$, these new absorption probabilities depend on six such quantities. For illustrative purposes, we will only consider $P_\infty ^{(m)}(1,0,0)$ as the others are computed via similar methods. We display the summation form of this probability:
$$P_\infty ^{(m)}(1,0,0)=\sum _{t=1}^\infty |\langle 0,L|Q\left(\Pi _\text{no}^nQ\right) ^{t-1}|m,R\rangle |^2.$$

We thus construct the following generating function:
$$r_\infty ^{(m)}(z)=\sum _{t=1}^\infty \langle 0,L|Q\left(\Pi _\text{no}^nQ\right) ^{t-1}|m,R\rangle z^t.$$
We define $l_\infty ^{(m)}(z)$ and $s_\infty ^{(m)}(z)$ similarly. The following proposition gives us a closed form for $r_\infty ^{(m)}(z)$:
\begin{proposition}
The generating function $r_\infty ^{(m)}(z)$ has the closed form:
\begin{align}
r_\infty ^{(m)}(z)=\left(\frac{3+2z+3z^2+(z-1)\sqrt{9+6z+9z^2}}{4z}\right)\left(\frac{-3-4z-3z^2+(z+1)\sqrt{9+6z+9z^2}}{2z}\right) ^{m-1}\
\end{align}
\end{proposition}
{\bf Proof:} Note that the aforementioned generating functions are all segmented by $|m-1\rangle |L\rangle$. By using the segmenting transformation in conjunction with induction, this leads to the conclusion:
$$r_\infty ^{(m)}(z)=r_\infty ^{(1)}(z)\left( l_\infty ^{(1)}(z)\right) ^{m-1}.$$
It remains to handle the $m=1$ generating functions. By combining a first step transformation and the previous equation, we have the system:
$$r_\infty ^{(1)}=\frac{2}{3}z+\frac{2}{3}zs_\infty ^{(1)}(z)-\frac{1}{3}zr_\infty ^{(1)}l_\infty ^{(1)}(z)$$
\begin{align}
s_\infty ^{(1)}=\frac{2}{3}z-\frac{1}{3}zs_\infty ^{(1)}(z)+\frac{2}{3}zr_\infty ^{(1)}l_\infty ^{(1)}(z)\
\end{align}
$$l_\infty ^{(1)}=-\frac{1}{3}z+\frac{2}{3}zs_\infty ^{(1)}(z)+\frac{2}{3}zr_\infty ^{(1)}l_\infty ^{(1)}(z).$$
Solving this system and choosing the root with a convergent Taylor series leads to the formula in the proposition. $\hfill\Box$

Recognizing that $P_\infty ^{(m)}(1,0,0)=\left( r_\infty ^{(m)}(z)\odot\overline{ r_\infty ^{(m)}(\bar{z})}\right) (1)$, it remains to compute the Hadamard product. For the $m=1$ case, we resort to direct integration as in Theorem 4.1.
\begin{theorem}
The following holds:
\begin{align}
P_\infty ^{(1)}(1,0,0)=\frac{5\sqrt{2}}{2\pi}-\frac{3\csc^{-1}(3)}{4\pi}-\frac{3}{8}\approx 0.6693.\
\end{align}
\end{theorem}

\subsection*{Finite Case}

We now compute the absorption probability $P_n^{(m)}(1,0,0)$ for the finite absorbing quantum walk $\Pi _\text{no}^n\Pi _\text{no}^0Q\leftrightarrow (\Z ,\tilde{C}_1,G_3,\{ 0,n\} )$. We omit proofs to encourage readability; refer to the previous section and Wang et. al. \cite{wang16} for details. As before, the right absorption probability may be written as:
$$P_n^{(m)}(1,0,0)=\sum _{t=1}^\infty |\langle 0,L|Q\left(\Pi _\text{no}^n\Pi _\text{no}^nQ\right) ^{t-1}|m,R\rangle |^2.$$
We thus construct generating functions $r_n^{(m)}(z)$ which have the form:
$$r_n^{(m)}=\sum _{t=1}^\infty \langle 0,L|Q\left(\Pi _\text{no}^n\Pi _\text{no}^nQ\right) ^{t-1}|m,R\rangle z^t.$$
The generating functions $l_n^{(m)}(z)$ and $s_n^{(m)}(z)$ are defined similarly. We compute closed form representations of these functions:
\begin{proposition}
The following holds:
\begin{align}
r_n^{(m)}(z)=2(z+1)(z-1)^{m-1}z^m\left(\frac{R_{n-m}(z)}{R_n(z)+z^2(1+3z)R_{n-1}(z)}\right)\
\end{align}
\begin{align}
l_n^{(m)}(z)=(z-1)^mz^m\left(\frac{R_{n-m}(z)+z^2(1+3z)R_{n-m-1}(z)}{R_n(z)+z^2(1+3z)R_{n-1}(z)}\right) .\
\end{align}
Here, $R_n(z)=\lambda _+(z)^n-\lambda _-(z)^n$ and \\ $\lambda _\pm (z)=\frac{z-1}{2}\left[ -(3z^2+4z+3)\pm\sqrt{(3z^2+4z+3)^2-4z^2}\right]$.
\end{proposition}

We now wish to compute Hadamard products of these generating functions. To do this, we will require a few lemmas.
\begin{lemma}
Let $f(w,z)=\frac{-z^2(1+3z)w+2z(z+1)}{-2z(z+1)w+(z+3)}$. Then $|w|,|z|\le 1\Rightarrow |f(w,z)|\le 1$.
\end{lemma}
\begin{lemma}
The following formulas hold:
\begin{align}
\overline{r_n^{(m)}\left(\frac{1}{\bar{z}}\right)}=-2(z+1)(z-1)^{m-1}z^m\left[\frac{R_{n-m}(z)}{R_n(z)-(z+3)R_{n-1}(z)}\right]\
\end{align}
\begin{align}
\overline{l_n^{(m)}\left(\frac{1}{z}\right)}=(1-z)^mz^m\left[\frac{R_{n-m}(z)-(z+3)R_{n-m-1}(z)}{R_n(z)-(z+3)R_{n-1}(z)}\right] .\
\end{align}
\end{lemma}
\begin{lemma}
The following formula holds:
$$\frac{R_{n-m}(z)}{(R_n(z)+z^2(1+3z)R_{n-1}(z))(R_n(z)-(z+3)R_{n-1}(z))}$$
\begin{align}
=\frac{1}{(3+z+z^2+3z^3)z^{2m-2}(z-1)^{2m-2}R_1(z)}\left[\frac{R_m(z)-(z+3)R_{m-1}(z)}{R_n(z)-(z+3)R_{n-1}(z)}-\frac{R_m(z)+z^2(1+3z)R_{m-1}(z)}{R_n(z)+z^2(1+3z)R_{n-1}(z)}\right] .\
\end{align}
\end{lemma}

We are now ready to compute the absorption probabilities via the Hadamard product.
\begin{theorem}
Let $\omega =\frac{1}{3}+\frac{2\sqrt{2}}{3}i$, $\delta _\pm (z)=\frac{1}{2}\left[ -(3z^2+4z+3)\pm\sqrt{(3z^2+4z+3)^2-4z^2}\right]$, $F_n(z)=\delta _+(z)^n-\delta _-(z)^n$, and $B_n(z)=\delta _+(z)^n+\delta _-(z)^n$. Then the right absorption probability satisfies the following:
\begin{align}
P_n^{(m)}(1,0,0)=\frac{1}{2}\left[\frac{F_{n-m}(1)F_{m-1}(1)}{\sqrt{6}F_{n-1}(1)}-\frac{F_{n-m}(\omega )B_{m-1}(\omega )}{\sqrt{2}B_{n-1}(\omega )}\right] .\
\end{align}
\end{theorem}

\subsection*{Discussion}

In this three state quantum walk we observe much of the same behavior as we did in the two state quantum walk. For instance, the semi-infinite walk is not recurrent; that is, $P_\infty ^{(1)}(1,0,0)<1$. Also, as with the two state quantum walk, the finite absorption probability does not limit to the semi-infinite absorption probability:
\begin{align}
P_\infty ^{(1)}(1,0,0)<\lim _{n\rightarrow\infty} P_n^{(1)}(1,0,0).\
\end{align}

It was proven in Wang et.\ al.\ \cite{wang16} that the finite absorption probabilities for $m=1$ satisfy the following recursion:
\begin{theorem}
The following formula holds:
\begin{align}
P_{n+1}^{(1)}(1,0,0)=\frac{2+3P_n^{(1)}(1,0,0)}{3+4P_n^{(1)}(1,0,0)}.\
\end{align}
\end{theorem}
However, a third order linear recursion in position will not hold for the absorption probabilities of these walks as the terms $F_{n-m}(\omega )B_{m-1}(\omega )$ and $F_{n-m}(1)F_{m-1}(1)$ separately satisfy third order recursions in $m$. We reconcile this observation in the following theorem:
\begin{theorem}
Let $p_m=P_n^{(m)}(1,0,0)$. Then the following recursion holds:
\begin{align}
p_{m+6}-134p_{m+5}+3599p_{m+4}-6932p_{m+3}+3599p_{m+2}-134p_{m+1}+p_m=0\
\end{align}
\end{theorem}

Absorption probability limits similar to Theorem 4.4 can be written, but are unwieldy. Additionally, it is clear from inspection that $P_\infty ^{(1)}(1,0,0)<\lim _{n\rightarrow\infty}P_n^{(1)}(1,0,0)$ as was the case for the two-state walks.

A numerical analysis of these absorption probabilities illustrates the effects of localization on this system. From figure 4, it appears that we have $P_n^{(m)}(1,0,0)+P_n^{(n-m)}(0,0,1)\le 1$ where equality only holds when $m=1$. This seems to support the idea that placing an absorbing boundary directly adjacent to an appropriately initialized particle will break down localization. 
\begin{figure}
\begin{center}\includegraphics[scale=0.9]{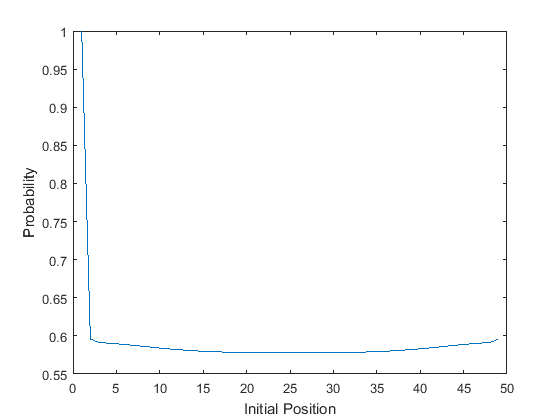}\end{center}
\caption{Plot of $P_n^{(m)}(1,0,0)+P_n^{(n-m)}(0,0,1)$ for $n=50$}
\end{figure}
Analytic expressions for the one-dimensional absorption probabilities of sections 4 and 5 are displayed in the charts of figure 5. If we let $a_n=P_n^{(1)}(1,0)$ referring to the $(\Z ,C_1,H)$ absorption probability of the previous section, and let $b_n=P_n^{(1)}(1,0,0)$ be the current Grover absorption probability, then we find $b_n=a_{2n-1}$. We can clearly see this by iterating the recursion in equation (43) and comparing to equation (55) (i.e. $P_{n+2}^{(1)}(1,0)=\frac{2+3P_n^{(1)}(1,0)}{3+4P_n^{(1)}(1,0)}$). Note also from figure 5 that for $m\ge 2$, the Grover absorption probabilities gain significant arithmetic complexity over the $m=1$ counterparts. This can be explained by equation (53) where the left term vanishes if $m=1$ (i.e. $F_0(1)=0$). Comparing this to the observation made of figure 4, we conclude that the term $\frac{F_{n-m}(1)F_{m-1}(1)}{\sqrt{6}F_{n-1}(1)}$ in equation (53) is the ''localization" component of the right absorption probabilities $P_n^{(m)}(1,0,0)$; that is, when $m=1$ there is no localization in the system and the term vanishes, and conversely when $m\ge 2$ localization is present and the term takes value.
\begin{figure}
\begin{center}\includegraphics[scale=0.65]{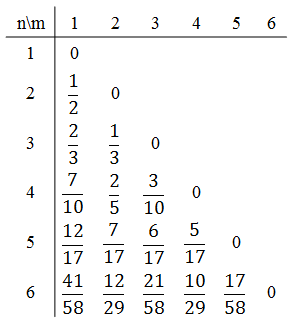}\hspace{2cm}\includegraphics[scale=0.65]{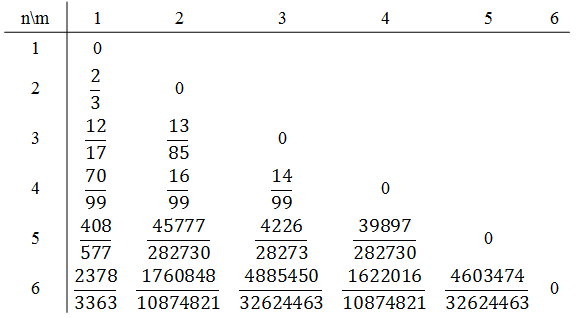}\end{center}
\caption{(\emph{Left}) Absorption probabilities $P_n^{(m)}(1,0)$ corresponding to the $(\Z ,C_1,H)$ quantum walk (\emph{Right}) Absorption probabilities $P_n^{(m)}(1,0,0)$ corresponding to the $(\Z ,\tilde{C}_1,G_3)$ quantum walk}
\end{figure}

\section{$(\Z ^d,C_d,G_{2d})$ Quantum Walk}

We now consider absorption probabilities for the $\Pi _\text{no}^{B_0}Q\leftrightarrow (\Z ^d,C_d,G_{2d},B_0)$ absorbing walk where $C_d=\{ S_{\pm 1},...,S_{\pm d}\}$ is the set of unit vectors in $\Z ^d$, $B_k=\{ z\in\Z ^d:\varphi _1(z)=k\}$ where $\varphi _k:\Z ^d\rightarrow\Z$ is a projection onto the $k^\text{th}$ coordinate, and $G_{2d}=\frac{1}{d}\begin{bmatrix}1 &\hdots & 1 \\ \vdots & \ddots & \vdots \\ 1 & \hdots & 1\end{bmatrix}-I_{2d}$ is the $2d\times 2d$ Grover matrix. We refer to $B_k$ as an \emph{absorbing wall}. We wish to calculate $P_\infty ^{(m)}$, or the probability that the quantum walk particle initialized in $|(m,0,...,0)\rangle |S_{+1}\rangle$ is eventually absorbed by $B$. Here, $|S_{+1}\rangle$ represents the internal state which translates perpendicular away from the absorbing boundary. We write the absorption probability as follows:
$$P_\infty ^{(m)}=\sum _{b\in B}\sum _{t=1}^\infty |\langle b,S_{-1}|Q(\Pi _\text{no}^{B_0}Q)^{t-1}|(m,0,...,0),S_{+1}\rangle |^2.$$
In this section, we will only illustrate computation of semi-infinite probabilities as the finite probabilities may be computed with similar methods as described previously.

In view of the summation form of the absorption probability, we construct the following generating functions:
$$f_{S_{\pm k},x}^{(m)}(z)=\sum _{t=1}^\infty\langle x,S_{-1}|Q(\Pi _\text{no}^{B_0}Q)^{t-1}|(m,0,...,0),S_{\pm k}\rangle z^t.$$
Here, $x\in B_0$. Solving closed form expressions for these generating functions is not as straightforward as it was in the one-dimensional walks, so instead of proving the closed form in a single result, we opt to elucidate the process through a series of lemmas and propositions.
\begin{lemma}
The following relation holds:
\begin{align}
f_{S_{\pm k},x}^{(m)}(z)=\sum _{y\in B_0}f_{S_{\pm k},y}^{(1)}(z)f_{S_{-1},x-y}^{(m-1)}(z).\
\end{align}
\end{lemma}
{\bf Proof:} The result follows from noting that $f_{S_{\pm k},x}^{(m)}(z)$ is segmented by $B_{m-1}$ and conducting a segmenting transformation. $\hfill\Box$
\begin{lemma}
The following system is satisfied:
\begin{align}
f_{S_{\pm k},x}^{(1)}(z)=\begin{cases} 
      \left(\frac{1}{d}-\delta _{S_{-1},S_{\pm k}}\right) z+\left(\frac{1}{d}-\delta _{S_{+1},S_{\pm k}}\right) z f_{S_{+1},x}^{(2)}(z) \\ +z\sum\limits_{S\ne S_{\pm 1}}\left(\frac{1}{d}-\delta _{S,S_{\pm k}}\right) f_{S,x-S}^{(1)}(z) & x=0 \\
      \left(\frac{1}{d}-\delta _{S_{+1},S_{\pm k}}\right) z f_{S_{+1},x}^{(2)}(z)+z\sum\limits_{S\ne S_{\pm 1}}\left(\frac{1}{d}-\delta _{S,S_{\pm k}}\right) f_{S,x-S}^{(1)}(z) & x\ne 0
   \end{cases}.\
\end{align}
\end{lemma}
{\bf Proof:} The result follows from a first step transformation and noting that if $x\ne 0$, the first step cannot be left. $\hfill\Box$

The system of generating functions described by lemma 6.2 is infinite for $d\ge 2$. As such we need some sort of construction to reduce this infinite system to a finite one. Let us introduce the functions $F_{S_{\pm k}}^{(m)}:\C\times\R ^d\rightarrow\C$ defined as follows:
$$F_{S_{\pm k}}^{(m)}(z,\Theta )=\sum _{x\in B_0}f_{S_{\pm k},x}^{(m)}(z)e^{ix\cdot\Theta} .$$
Since $\varphi _1(x)=0$ for $x\in B_0$, we can informally think of $\Theta\in\R ^{d-1}$ as opposed to $\R ^d$. These new functions allow us to rewrite the previous lemmas to result in a finite system. We first reformulate the segmenting relation from lemma 6.1.
$$F_{S_{\pm k}}^{(m)}(z,\Theta )=F_{S_{\pm k}}^{(1)}(z,\Theta )\left( F_{S_{-1}}^{(1)}(z,\Theta )\right) ^{m-1}.$$
We may now write the following system of $2d$ functions:
\begin{align*}
F_{S_{\pm k}}^{(1)}(z,\Theta ) &= \left(\frac{1}{d}-\delta _{S_{-1},S_{\pm k}}\right) z+\left(\frac{1}{d}-\delta _{S_{+1},S_{\pm k}}\right) z F_{S_{+1}}^{(1)}(z,\Theta)F_{S_{-1}}^{(1)}(z,\Theta ) \\
	&+ z\sum _{S\ne S_{\pm 1}}\left(\frac{1}{d}-\delta _{S,S_{\pm k}}\right) e^{-iS\cdot\Theta}F_{S}^{(1)}(z,\Theta ) .
\end{align*}
Upon solving this system, we can take an inverse Fourier transform of these new generating functions to get back to the $f_{S_{\pm k},x}^{(m)}(z)$ generating functions:
$$f_{S_{\pm k},x}^{(m)}(z)=\frac{1}{(2\pi )^d}\int _{\lVert\Theta\rVert _\infty\le\pi}F_{S_{\pm k}}^{(m)}(z,\Theta )e^{-ix\cdot\Theta}d\Theta$$
Once we get back the $f$ generating functions, we may compute the absorption probability by summing over the usual Hadamard products:
$$P_\infty ^{(m)}=\sum _{x\in B_0}\left( f_{S_{+1,x}}^{(m)}(z)\odot\overline{f_{S_{+1,x}}^{(m)}(\bar{z})}\right) (1)$$
Often, the inverse Fourier transform and the summation are challenging to analytically compute.

Alternatively, if we let $\mathbf{x}=e^{i\Theta}$ and $G_{S_{\pm k}}^{(m)}(z,{\bf x} )=F_{S_{\pm k}}^{(m)}(z,\Theta )$, then $G_{S_{\pm k}}^{(m)}(z,{\bf x} )$ is a power series in $z$ and a multivariate Laurent series in $\mathbf{x}$. To extract the absorption probability from this generating function, we need an extension of the Hadamard product which handles Laurent series of several complex variables \cite{krantz01}. If we let $d=2$ and follow the proof of proposition 2.5, the integral representation of this absorption probability would ostensibly take the form
$$P^{(m)}_\infty =-\frac{1}{4\pi ^2}\iint _\Gamma\frac{1}{xz}G_{S_{+1}}^{(m)}(z,x)\overline{G_{S_{+1}}^{(m)}\left(\frac{1}{\bar{z}},\frac{1}{\bar{x}}\right)}dxdz$$
for some surface $\Gamma\subset\C ^2$. However, the development of the Hadamard product is insufficiently merged with the theory of complex analysis with several variables to properly determine what this region of integration should be. Regardless, we can still write closed form expressions for these generating functions for $d=2$. In particular, we find:
$$F_{S_{+1}}^{(1)}(z,\theta)=\frac{2(z^4+(z^3+z)\cos\theta +1)\pm 2\sqrt{(z^2-1)^2(z^2+z(\cos\theta -1)+1)(z^2+z(\cos\theta +1)+1)}}{z(z^2+2z\cos\theta +1)}$$
Based on numerical results, this semi-infinite absorption probability is approximately $P_\infty ^{(1)}\approx 0.646$.

We can also write a generating function recursion relating to the $\Pi _\text{no}^{B_n}\Pi _\text{no}^{B_0}Q\leftrightarrow (\Z ^d,C_d,G_{2d},B_0\cup B_n)$ double absorbing wall quantum walk. In this case, we wish to find the probability that the particle is absorbed by $B_0$ before it is absorbed at $B_n$. Let us write this absorption probability as:
$$P_n ^{(m)}=\sum _{b\in B_0}\sum _{t=1}^\infty |\langle b,S_{-1}|Q(\Pi _\text{no}^{B_n}\Pi _\text{no}^{B_0}Q)^{t-1}|(m,0,...,0),S_{+1}\rangle |^2.$$
If we write the generating functions
$$F_{S_{\pm k},n}^{(m)}(z,\Theta )=\sum _{x=-\infty}^\infty\sum _{t=1}^\infty\langle x,S_{-1}|Q(\Pi _\text{no}^{B_n}\Pi _\text{no}^{B_0}Q)^{t-1}|(m,0,...,0),S_{\pm k}\rangle z^te^{ix\cdot\Theta}$$
then we can use the methods relating to the single absorbing wall problem to compute this absorption probability. For $d=2$, these generating functions satisfy the recursion:
$$F_{S_{\pm k},n+1}^{(1)}(z,\theta )=z\left(\frac{z^2(\cos\theta +z)F_{S_{\pm k},n}^{(1)}(z,\theta )-(z^2+2z\cos\theta +1)}{z(z^2+2z\cos\theta +1)F_{S_{\pm k},n}^{(1)}(z,\theta )-(\cos\theta +z)}\right)$$
Lemma 2.1 may be used to derive a closed form for these generating functions. Though the analytic results are currently out of reach, based on the numerical data we present a conjecture.
\begin{conjecture}
For $d=2$, we have $\lim _{n\rightarrow\infty}P_n^{(1)}=\frac{2}{3}$.
\end{conjecture}
Unfortunately, the numerical data does not support a linear fractional recursion in $n$ governing the absorption probabilities, as was the case for the one dimensional quantum walks. However, from the numerical data it seems clear that $P_\infty ^{(1)}<\lim _{n\rightarrow\infty}P_n^{(1)}$.

\section{Conclusion}

In this paper we have developed absorption probabilities for two and three state one dimensional quantum walks, and made partial progress toward computing absorption probabilities of higher dimensional quantum walks with absorbing walls. In these examples we have found that the presence of an absorbing boundary far from the initial position of the walker paradoxically increases the probability that the particle is absorbed at a near boundary. We resolve this phenomenon by considering that the quantum walk exhibits wavelike behavior and that information is reflected by an absorbing boundary. For the one dimensional walks, we have found two recursions which govern these absorption probabilities in both initial condition and boundary placement.

It is the hope of the authors that these techniques may be extended to more general absorption settings. In particular, it would be interesting to compute absorption probabilities for a quantum walk in a $d$-dimensional absorbing box, or for a $d$-dimensional quantum walk with a single absorption unit at the origin. These results would require more sophisticated path counting arguments than the ones described above, but the procedure for computing the Hadamard product for the resulting generating functions would likely remain the same. It would also be interesting to extend these arguments to absorption probabilities for electric quantum walks \cite{oka05}\cite{cedzich13}.

\bibliography{biblio}
\bibliographystyle{plain}

\end{document}